\documentclass[10pt,conference]{IEEEtran}
\IEEEoverridecommandlockouts
\usepackage{tikz}
\usepackage{balance}
\usetikzlibrary{shapes.geometric, arrows, positioning}
\usepackage{cite}
\usepackage{amsmath,amssymb,amsfonts}
\usepackage{graphicx}
\usepackage{textcomp}
\usepackage{algpseudocode}
\usepackage{listings} 
\usepackage[most]{tcolorbox}
\usepackage{booktabs} 
\usepackage{xcolor} 

\ifdefined\drawcomments
  \newcommand{\yuki}[1]{\textcolor{purple}{[Yuki: #1]}}
  \newcommand{\lili}[1]{\textcolor{cyan}{[Lili: #1]}}
  \newcommand{\rufeng}[1]{\textcolor{orange}{[Rufeng: #1]}}
  \newcommand{\zixu}[1]{\textcolor{blue}{[Zixu: #1]}}
\else
  \newcommand{\yuki}[1]{\ignorespaces}
  \newcommand{\lili}[1]{\ignorespaces}
  \newcommand{\rufeng}[1]{\ignorespaces}
  \newcommand{\zixu}[1]{\ignorespaces}
\fi

\newcommand{\blacknumber}[1]{%
  \tikz[baseline=(char.base)]{
    \node[shape=rectangle,fill=black, text=white, font=\sffamily\bfseries, inner sep=2pt] (char) {#1};
  }
}

\usepackage[hidelinks]{hyperref}

\definecolor{codegreen}{rgb}{0,0.6,0}
\definecolor{codegray}{rgb}{0.5,0.5,0.5}
\definecolor{codepurple}{rgb}{0.58,0,0.82}
\definecolor{backcolour}{rgb}{0.95,0.95,0.92}

\lstdefinestyle{style_java_1}{
    backgroundcolor=\color{backcolour},   
    commentstyle=\color{codegreen},
    keywordstyle=\color{magenta},
    numberstyle=\tiny\color{codegray},
    stringstyle=\color{codepurple},
    basicstyle=\ttfamily\small,
    breakatwhitespace=false,         
    breaklines=true,                 
    captionpos=b,                    
    keepspaces=true,                 
    numbers=left,                    
    numbersep=5pt,                  
    showspaces=false,                
    showstringspaces=false,
    showtabs=false,                  
    tabsize=2,
    frame=ltb,
    framerule=0pt,
}
\usepackage{url}  

\begin{document}

\title{Characterizing Bugs in Login Processes of Android Applications: An Empirical Study\\
\thanks{Yepang Liu is affiliated with both the Research Institute of Trustworthy Autonomous Systems and Department of Computer Science and Engineering.}
}
\author{
    \IEEEauthorblockN{ Zixu Zhou}
    \IEEEauthorblockA{
    \textit{McGill University}\\
    Montreal, Canada \\
    zixu.zhou@mail.mcgill.ca}
    \and
    \IEEEauthorblockN{Rufeng Chen}
    \IEEEauthorblockA{
    \textit{McGill University}\\
    Montreal, Canada \\
    rufeng.chen@mail.mcgill.ca}
    \and
    \IEEEauthorblockN{ Junfeng Chen}
    \IEEEauthorblockA{ 
    \textit{Southern University of Science and Technology}\\
    Shenzhen, China \\
    chenjf@mail.sustech.edu.cn}
    \and
    \IEEEauthorblockN{Yepang Liu}
    \IEEEauthorblockA{
    \textit{Southern University of Science and Technology}\\
    Shenzhen, China \\
    liuyp1@sustech.edu.cn}
    \and
    \IEEEauthorblockN{Lili Wei}
    \IEEEauthorblockA{
    \textit{McGill University}\\
    Montreal, Canada \\
    lili.wei@mcgill.ca}
}

\maketitle
\begin{abstract}
The login functionality, being the gateway to app usage, plays a critical role in both user experience and application security. As Android apps increasingly incorporate login functionalities, they support a variety of authentication methods with complicated login processes, catering to personalized user experiences.
However, the complexities in managing different operations in login processes make it difficult for developers to handle them correctly.
In this paper, we present the first empirical study of login issues in Android apps. We analyze 361 issues from 44 popular open-source Android repositories, examining the root causes, symptoms, and trigger conditions of these issues. Our findings indicate that the vast majority of the login issues are induced by the improper handling of complex state transitions during the login process, which can prevent users from logging in or misdirect them to incorrect subsequent actions. Additionally, we observed that issues related to this cause typically require the convergence of multiple trigger conditions to manifest. These findings can help developers to model the login processes which can help them to identify the causes of issues and design targeted test cases and precise test oracles. 
Our dataset has been made openly available to facilitate future research in this area.

\end{abstract}


\section{Introduction}
Login plays a critical role in Android apps, typically representing the first point of interaction between the application and the users~\cite{Alsanousi2023}.
Issues in login processes can significantly degrade the user experience~\cite{FTUE,5reasonsfailuser,Mobileappusage,9042272}.
Studies have shown that first-time login failures can lead to a direct and immediate drop in user engagement by as much as 25\%~\cite{FTUE,10679403}. Such statistics highlight the critical nature of robust login processes and demonstrate their potential economic impact on software developers, emphasizing the need for improved reliability in the login processes.

 Existing studies have extensively explored a wide spectrum of software defects and usability challenges in Android applications~\cite{Alsanousi2023,10679403,10.1145/3386685}. Studies were conducted to characterize diverse issues in the Android ecosystem ranging from compatibility issues~\cite{huang2023conffix,mahmud2023detecting,yang2023compatibility,chen2024demystifyingdevicespecificcompatibilityissues}, WebView problems~\cite{luo2011attacks,8999997,hu2023omegatest}, cross-platform usability challenges~\cite{steinbock2024comparing,chen2024your,yu2021layout} to specific security vulnerabilities~\cite{10.1145/3545948.3545955,10504267,7589802,10.1145/3301285,10.1145/3321705.3329831,10738442,7962350}.
 Some related studies analyze vulnerabilities related to Android app authentication, an important component in the login process~\cite{10.1145/3551349.3559524,10628996,10.1145/3321705.3329801,8952200}.
 For example, Philippaerts et al.~\cite{10.1145/3551349.3559524} proposed Oauch, which can measure how well individual identity providers (IdPs) implement the security specifications defined in the OAuth standard and provide development suggestions to developers.
 Jannett et al.~\cite{10628996} conducted an empirical study on Single Sign-On (SSO) and developed a tool, Sok, focusing on testing SSO in Android by large-scale dual-window analysis. 
 Shi et al.~\cite{10.1145/3321705.3329801} developed MoSSOT, a backbox tester for the OAuth process in Android. Furthermore, Tamjid et al.~\cite{8952200} conducted an empirical study on the usage of OAuth APIs and their implications for mobile security, leading to the development of OAUTHLINT, a tool to identify vulnerabilities in OAuth implementations for Android apps. These studies mainly focused on vulnerabilities arising from the authentication process.
 However, the login process extends beyond authentication.

The login processes can be briefly described in three stages: Pre-login, login, and post-login. Initially, in the Pre-Login phase, the app checks for an existing valid token to either bypass or prompt for user credentials. During the Login phase, the app assesses the credentials and, if necessary, proceeds with multi-factor authentication before granting access. Once authentication is successful, the Post-Login phase commences, granting the user full access to the app's features.
 Developing reliable and robust login processes requires the app developers to properly manage complicated interactions between the app, the users and the servers.
 In addition, such interactions can interfere with the Android system events.
 Such complexities make it especially challenging to implement the login processes correctly.

No existing studies have systematically characterized the various issues in the login processes of Android apps. This paper presents an empirical analysis of the root causes, symptoms, and trigger conditions of login issues in Android apps to address this research gap. 
Our investigation is structured around the following research questions:
\begin{itemize}
    \item \textbf{RQ1: (Root Causes of Login Issues):}  \textit{What are the common root causes of login issues in Android apps?}
    
    \item \textbf{RQ2: (Symptoms of Login Issues):} \textit{What symptoms are commonly observed when login issues occur, and how do they impact user experience?}
    
    \item \textbf{RQ3: (Trigger Conditions of Login Issues):} \textit{Under what conditions are login issues likely triggered?}
\end{itemize}

Answering these research questions offers practical benefits to both developers and researchers.
RQ1 characterizes the common root causes of login issues in Android apps to understand how the issues are induced comprehensively. RQ2 and RQ3 investigate the symptoms and trigger conditions of login issues and can guide developers to generate targeted test cases and effective test oracles.

We analyzed 361 login issues across 44 open-source Android apps to answer our research questions. 
This analysis leads to several key findings.
We proposed using state machines to model the complicated login processes. We disclosed that the majority (62\%) of the login issues can be attributed to errors in managing the states and their transitions.
For example, Home Assistant~\cite{home-assistant-android} requires users to input a URL for login to access remote services. The issue~\cite{HomeAssistantIssue} arose because developers inadequately managed the login process's state; they did not properly \lili{promptly or properly?}\zixu{Done}store and update the new URL provided by the user, leading to a blank screen even with the correct credentials entered. 
This observation suggests that future research can focus on addressing such login flow errors.
Our proposed state machine can guide the test generation for exposing these errors.
In addition, our study reveals that login flow errors can induce various consequences, including failures in the authentication process, crashes, etc.
These observations can help define the test oracles for effective login flow error detection.

To summarize, this paper makes the following major contributions:
\begin{itemize}
\item \textbf{Login issues dataset:} We carried out the first empirical study on login issues within real-world Android apps. Our dataset includes 361 login issues sourced from 44 popular repositories. \textit{We have made this dataset available to support further research}~\cite{code}.

\item \textbf{Comprehensive Categorization:} We systematically identified and categorized the root causes, symptoms, and trigger conditions of login issues in Android apps.
\item \textbf{Bridging Theory and Practice:} By exploring the implications of our findings, this research not only can aid developers in modeling the login process but also help them to design targeted test cases and precise test oracles.
\end{itemize}

All datasets and scripts used in this study are open-sourced and available, further enabling the community to replicate our findings and extend our work~\cite{code}.


\section{Background}
\subsection{Android Authentication Methods}
Android apps utilize various authentication methods to manage user access and ensure security. Authentication is a critical component of the login process, serving as the method by which applications verify user identity to grant access to their features and ensure security.
\lili{One small gap here is: what are the relationships between authentication and login?}\zixu{Added a sentence to explain auth belongs to login}
The most common approaches include traditional username and password login, social media integration, and biometrics such as fingerprint scanning and facial recognition~\cite{9068715}. Each method offers distinct advantages and disadvantages. For instance, username and password combinations are widely used due to their simplicity and familiarity but are vulnerable to brute force attacks and phishing. Social media logins such as Google Lgoin~\cite{google-login-in} and Facebook Login~\cite{facebook-login-in} offer convenience by allowing users to sign in with existing accounts, reducing password fatigue and streamlining the user experience. However, they can raise concerns about privacy and data security. Biometric methods provide robust security and a quick authentication process but require specialized hardware and can be affected by environmental factors or changes in the user's physical condition. To further enhance security, Multi-Factor Authentication (MFA)~\cite{10348624} is increasingly being implemented, which requires users to provide multiple verification factors, significantly reducing the risk of unauthorized access.

\subsection{States in the Login Process}
The login process in Android apps can be divided into three distinct phases: pre-login, login, and post-login. 

\textbf{Pre-login Phase:} This initial stage involves checking whether the user has previously logged in and if there is a valid token present. If no valid token is found, the user is prompted to select a login approach and enter the corresponding credentials. This stage is crucial for determining the starting point of the authentication process, ensuring that returning users can proceed faster if they have valid sessions.

\textbf{Login Phase:} During this phase, the submitted credentials are validated to verify correctness. Additionally, if Multi-Factor Authentication (MFA) is enabled, the user is required to provide an MFA code. This code is then validated to ensure an extra layer of security. This phase is critical as it directly impacts the security of the user's access and the integrity of the application.

\textbf{Post-login Phase:} After successful verification, this phase grants the user access to the app's resources. It marks the transition from authentication to actual usage of the application, where the user interacts with the app's features based on their access level and permissions. This phase is essential for a seamless transition and ensuring the user's experience is smooth and secure.

\section{emprical study setup}
A thorough understanding and categorization of these problems is essential to address the pervasive login issues in Android apps. We followed the process adopted by existing work \cite{8999997,chen2024demystifyingdevicespecificcompatibilityissues,10.1145/3597926.3598138,7582761} to prepare the dataset for our empirical study. Our research began with an extensive examination of Android-related repositories on GitHub \cite{github} using specific keywords such as ``android-app, ``android" and ``android-library," among others, to select repositories likely to contain relevant data on login issues. All keywords used are available on our GitHub repository~\cite{code}.

\subsection{Data Collection}

Our study involved a meticulous selection process for identifying repositories that are both relevant and exhibit high quality based on the following criteria:
1) \textbf{Recent Activity:} Repositories must have had their last commit after January 1, 2023, indicating active development;
2) \textbf{Community Engagement:} Repositories were required to have more than 50 stars, reflecting a certain level of community endorsement;
3) \textbf{Substantial Contribution:} Each repository should have over 200 commits, demonstrating significant developer investment;
4) \textbf{Programming Language:} Repositories needed to predominantly use Java or Kotlin, which are commonly used for Android development.

\begin{figure}[h]
    \centering
    \begin{tikzpicture}[node distance=0.5cm, auto]
        \tikzset{
            startstop/.style={rectangle, rounded corners, minimum width=0.5\linewidth, minimum height=0.4cm, text centered, draw=black, fill=red!10, font=\small},
            result/.style={rectangle, minimum width=0.5\linewidth, minimum height=0.4cm, text centered, draw=black, fill=blue!10, font=\small},
            process/.style={rectangle, minimum width=0.8\linewidth, minimum height=0.4cm, text centered, draw=black, fill=orange!10, font=\small},
            arrow/.style={thick,->,>=stealth, shorten >=1pt, shorten <=1pt},
            description/.style={text width=5cm, align=left, midway, font=\small} 
        }

        \node (start) [startstop] {GitHub};
        \node (in1) [result, below=of start] {2675 Repositories};
        \node (in2) [result, below=of in1] {50201 Issues From 2569 Repositories};
        \node (in3) [result, below=of in2] {2398 Issues From 89 Repositories};
        \node (out1) [result, below=of in3] {361 Issues From 44 Repositories};

        \draw [arrow] (start) -- (in1) node[description] {Step 1:GitHub Search};
        \draw [arrow] (in1) -- (in2) node[description] {Step 2: Keywords Filter};
        \draw [arrow] (in2) -- (in3) node[description] {Step 3: TF-IDF Keywords Filter };
        \draw [arrow] (in3) -- (out1) node[description] {Step 4: Advanced Criteria Filter };
    \end{tikzpicture}
     \caption{The process of dataset collection}
\label{fig:data selection}
\end{figure}
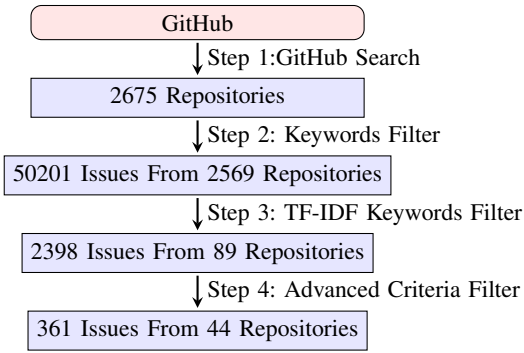

To efficiently gather data, we utilized the GitHub Search API \cite{githubsearch}, which imposes a cap of 1,000 results per query. The query example is shown in Listing~\ref{lst:query}. To circumvent this limitation and ensure comprehensive data retrieval, we adopted an iterative querying approach\cite{8595172,chen2024demystifyingdevicespecificcompatibilityissues,10148722}. By sorting repositories by their most recent update and progressively refining our search with the `pushed' filter, we were able to compile a complete list of repositories fitting our criteria. This methodological rigor enabled us to amass a total of 2,675 repositories. Fig \ref{fig:data selection} shows the overview of the data collection and selection process.  

The repositories' metadata and the complete list of repositories have been preserved and are accessible within our study's digital artifacts~\cite{code}. 
\begin{lstlisting}[language=SQL, caption={GitHub Search Query Example}, label=lst:query]
query := keyword
            + " stars:>50" 
            + " pushed:>=2023-01-01"
            + " language:java language:kotlin"
\end{lstlisting}

In the subsequent phase, our objective was to pinpoint login-specific issues within the amassed repositories. Initially, to refine our search and reduce extraneous data, we excluded repositories with names suggesting a focus on coding interviews or algorithmic challenges, such as those containing ``interview" or ``leetcode". After the filter, we obtained 2,569 repositories. 

To enhance search performance and avoid triggering rate limits of the GitHub search API, we downloaded all closed issues from 2,569 repositories. We then conducted a keyword analysis using TF-IDF \cite{aizawa2003information} and stopwords \cite{stopwords} to identify and extract key terms frequently appearing in login-related issues. Initially, we manually selected 10 login-related issues and used TF-IDF in conjunction with stopwords to determine the top 10 most frequent terms within these issues. These keywords were then employed to search through all downloaded issues, focusing on the titles and descriptions. Subsequently, we applied TF-IDF analysis again to the issues that contained these top 10 terms, such as ``login". To keep the login relevance of results issues, we only extracted the top 100 terms, such as ``auth", ``sign in" and ``MFA" that were used as search keywords. We manually verified each term to ensure the relevance of the login process. This refined list of keywords guided our secondary search through the GitHub issues API, leading us to identify 2,398 issues across 189 repositories.

To ensure the scientific validity and reliability of our study, we applied rigorous filters to the identified issues: 1) \textbf{Fix Commit Link:} Each issue had to include a link to a commit that purportedly resolved the issue, ensuring that we could verify and analyze the resolution approaches; 2) \textbf{Accessibility of Commit Links:} The commit links associated with the issues had to be accessible, allowing us to review the actual code changes made.

After applying these filters, our final dataset comprised 361 issues from 44 repositories. As detailed in Table~\ref{tab:android-issues-significant}, which only contains the repositories that have more than 10 issues, our dataset includes highly popular Android apps, evidenced by significant GitHub stars and download numbers on Google Play. This diverse set encompasses a wide range of app categories, ensuring a comprehensive analysis across various user experiences and functionalities. For instance, WordPress~\cite{wordpress} serves as a content management app, NextCloud~\cite{nextcloud-android} is utilized for file synchronization and cloud storage, and Element~\cite{element-android} functions as a secure messaging platform. This breadth of app types, from productivity to social interaction, enhances the relevance and applicability of our findings to real-world usage scenarios.

\begin{table}
  \caption{Summary of Android Open-Source Repositories with Significant Login Issues(We only listed repositories that contain more than 10 issues) \lili{The number of issues in this table adds up to 258. Is it correct?}\zixu{added caption} 
  }
  \label{tab:android-issues-significant}
  \centering
  \begin{tabular}{lccc}
    \toprule
    Repository & Issues & Stars& Downloads\\
    \midrule
    thunderbird/thunderbird-android \cite{thunderbird-android} & 41 & 10.9k & 50k+ \\
    woocommerce/woocommerce-android \cite{woocommerce-android} & 37 & 277 & 1M+\\
    nextcloud/android \cite{nextcloud-android} & 35 & 4.3k &1M+ \\
    firebase/FirebaseUI-Android  \cite{firebaseui-android}& 25 & 4.6k & N/A \\
    fossasia/open-event-attendee-android \cite{fossasia} & 17 &2k& 500k+ \\
    element-hq/element-android \cite{element-android} & 16 & 3.4k & 1M+ \\
    home-assistant/android \cite{home-assistant-android}& 16 &2.3k & 1M+ \\
    tuskyapp/Tusky \cite{tusky}& 15 &2.5k &500k+ \\
    commons-app/apps-android-commons \cite{commons-android-commons} & 14 &1k &100k+ \\
    ankidroid/Anki-Android \cite{anki-android}& 11 & 8.7k &10M+ \\
    fossasia/phimpme-android \cite{phimpme-android}& 11 & 2.6k & 1M+ \\
    element-hq/element-x-android \cite{element-x-android}& 10 & 1.1k & 10k+ \\
    bitfireAT/davx5-ose \cite{davx5-ose} & 10 & 1.5k & 100k+ \\
    \bottomrule
  \end{tabular}
\end{table}

\subsection{Data Analysis}

 The study observes that precise categorizations of issues play a vital role in the development of Android apps~\cite{10.1145/3386685}. We employed an open coding methodology to address the classification of 361 identified login issues into root causes, symptoms, and trigger conditions. This approach facilitated a systematic and structured classification process, enhancing the depth and accuracy of our analysis. 



\subsubsection{Stage I Initial Categorization}

The first stage involved a preliminary analysis where 10\% of the issues, totaling 36 distinct issues, were randomly selected for initial coding. Two researchers independently reviewed each issue’s description, developer replies, and fix commits. The objective was to identify broad categories under the three aspects: (1) \textbf{Root Causes}: Fundamental reasons that initiate the login issues; (2) \textbf{Symptoms}: Observable effects or behaviors resulting from the issues; (3) \textbf{Trigger Conditions}: Specific scenarios or conditions under which the issues occur. A combination of several conditions can trigger one issue. So, one issue can only have one label for the root cause and symptom but multiple for trigger conditions. 

Each researcher independently assigned issues to preliminary categories, ensuring that the initial classification captured the diversity and complexity of the login issues.

Following the independent phase, the two authors convened to discuss their findings and reconcile any differences in category assignment. A third author resolved any conflicts between the initial two authors. 

\subsubsection{Stage II Extensive Categorization and Category Refinement}

With the foundational categories established, the second stage extended the categorization process to an additional 30\% of the issues. This larger sample allowed for a robust application of the categorization framework. The two authors continued to work independently, applying the categories to new issues and meeting regularly to discuss their findings. These discussions allowed for continuous refinement of the categories: \textbf{(1)} Identifying new categories that emerged from the data; \textbf{(2)} Modifying existing categories to reflect the data better; \textbf{(3)} Ensuring consistency and accuracy in the application of categories across all analyzed data.

This iterative categorization and discussion cycle was repeated until all issues in the sample had been classified. The open coding procedure facilitated a dynamic adaptation of the categorization framework, accommodating the complexities and nuances of the login issues encountered.

\subsubsection{Final Review and Consensus}

The final stage of the open coding process involved a comprehensive review and consensus discussion to confirm the accuracy of category assignments and finalize the taxonomy of login issues. The categorization's reliability was highlighted by high agreement levels among researchers, quantified by Cohen’s Kappa: 0.88 for Root Causes, 0.82 for Symptoms, and 0.55 for Trigger Conditions. The relatively lower Kappa for Trigger Conditions is attributed to the complexity and variability in identifying multiple conditions per issue, where exact agreement on the type and number of conditions is required, reflecting substantial inter-rater reliability.  
\section{Empirical Study}
This section presents the findings of our empirical analysis of the 361 login issues.

\subsection{RQ1:Root Causes}
Understanding root causes is vital as it lays the groundwork for developing a more robust login process in Android apps.
Through a meticulous process of open coding, our study categorized these root causes into five groups, each characterized by distinct patterns and contributing factors. Table~\ref{tab:root-causes} shows an overview of all the categories.
\begin{table}
  \centering
  \caption{Summary of Root Causes and Issue Counts \lili{Update the table with the new names of each category}\zixu{Done}}
  \label{tab:root-causes}
   \begin{tabular}{@{}c l r@{}}
    \toprule
    \textbf{No.} & \textbf{Root Causes }& \multicolumn{1}{c}{\textbf{\#Issues}} \\
    \midrule
    1 & Login Flow Error & 214 \\ 
     & 1.1 Improper Error State Handling & 57  \\ 
     & 1.2 Interference with Other State Machines & 57   \\
     & 1.3 Missing Login Flow & 52 \\
     & 1.4 Incorrect Login Flow & 48  \\
    2 & API Misuse & 63  \\
    3 & Null Object Check Omission & 39 \\
    4 & Improper Encoding Algorithm & 23 \\
    5 & Wrong Environment Dependency & 22 \\
    \bottomrule
  \end{tabular}\\
\end{table}

\subsubsection{Login Flow Errors}
\begin{figure*}[ht!]
  \centering
  \includegraphics[width=0.9\linewidth]{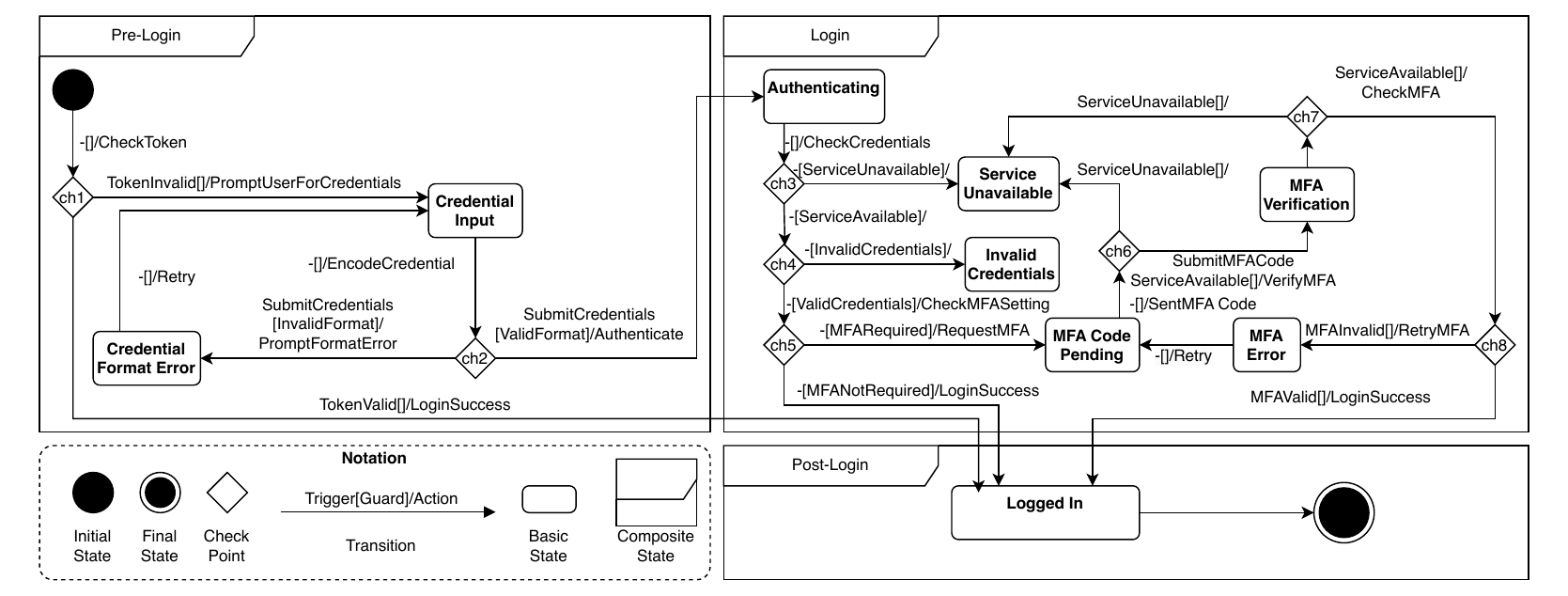}
  \caption{State Machine for Login Process}
  \label{fig:statemachine}
\end{figure*}
The login processes in Android apps typically require a dedicated sequence of operations (the login flow) involving the apps, the users, and some backend services. These operations can include user inputs, authentication, or error handling. For example, checking for the presence of a valid token will decide whether to transfer apps into logged-in or ask the user to input credentials. Subsequently, suppose the user inputs the correct credentials. In that case, the app will prompt the user for MFA processes or log in directly to the user, depending on whether the user activates MFA. 

We propose to leverage state machines to model this complicated process.
Figure~\ref{fig:statemachine} illustrates our proposed state machine, which is structured into three \textbf{composite states}: \textbf{Pre-Login}, \textbf{Login}, and \textbf{Post-Login}. The format denotes each transition between states: \textit{Trigger[Guard]/Action}, signifying that if the condition specified in the \textit{Guard} is met following a \textit{Trigger}, the app executes the corresponding \textit{Action}. Initially, the app checks for a valid token from the \textit{Initial State} within the \textbf{Pre-Login} phase; if a valid token is present, it transitions directly to the \textit{Logged In} state, thus bypassing the credential input. The app transfers to the \textit{Credential Input} state if no valid token exists, prompting the user to input login credentials. Subsequently, if the credentials are incorrectly formatted, the state transitions to \textit{Credential Format Error}; otherwise, the app proceeds to authenticate in the \textit{Authenticating} state under the \textbf{Login} phase. During authentication, the app moves to the \textit{Service Unavailable} state if the service is unavailable. Upon successful credential verification, and if multi-factor authentication (MFA) is required, the app transitions to the \textit{MFA Code Pending} state; if MFA is not required, or once MFA verification is complete without errors, the app then transitions to the \textit{Logged In} state. However, if MFA verification fails, the app enters the \textit{MFA Error} state. The \textbf{Post-Login} phase consists solely of the \textit{Logged In} state, indicating successful access to the app.

We proposed this state machine based on analyzing the login processes defined in our 44 analyzed Android apps. We initially examined the login methods supported by each app, discovering that 38 repositories utilize username and password authentication, and 18 of these also support MFA.
We downloaded all available repositories from Google Play to derive this state machine. We performed login actions for each app on an Android emulator using Appium\cite{appium}, a tool designed to facilitate UI automation that helps identify the current activity and corresponding method of an app. Subsequent analysis of the relevant source code revealed the complete process and the methods involved. We then abstracted the states and their transitions based on transitions between different methods, the conditions triggering these method calls, and the subsequent action.
While different apps can have variations in the login process (e.g. NextCloud\cite{nextcloud-android} allows users to use their domain name to login), our state machine captures the general login processes of the Android apps.
Login flow errors refer to improper handling of states or transitions in the login flow. They were divided into four subcategories, each reflecting challenges in maintaining the integrity of the login process through proper state transitions and error handling.
\paragraph{Improper Error State Handling}
Improper Error State Handling induced 57 issues where the app did not correctly handle errors that arose in the login process. An example is issue \#7437 from WordPress~\cite{wordpress}, where users were incorrectly informed that their email was not registered when a network error occurred during the login process. This problem was addressed in a subsequent pull request, correcting the behavior by displaying a generic network error message instead.

\paragraph{Interference with Other State Machines}
\lili{After a second thought, I changed the name of the category. Please be reminded to revise other places.}\zixu{Modified}
Android apps are fundamentally built around the concept of lifecycles, which manage how activities within the app are created, paused, and destroyed based on user interactions and system events. The Android lifecycle can be viewed as a state machine~\cite{google-android-doc-state}, where each lifecycle state represents a node, and transitions between these states are triggered by lifecycle events.
The login flow can easily interfere with the lifecycle events as it involves intensive transitions between Activities. 
57 issues in our dataset were induced by failures to consider such interference.
For example, in Thunderbird~\cite{thunderbird-android} and Wikimedia Commons Android apps~\cite{commons-android-commons}, users frequently encounter data loss issues when rotating their devices. Specifically, in Thunderbird issue \#4936 and Wikimedia Commons issue \#3973, users lose selected account information or entered credentials because the apps do not save the user input data during the lifecycle state transitions.
When the device is rotated, the login activity will be destroyed and then recreated.
The data will be lost if the user input data are not stored via \texttt{onSaveInstanceState()}.

\paragraph{Missing Login Flow}\label{Missingloginflow}
This category represented 52 \lili{Have you re-do the categorization for these 61 issues?}\zixu{Yes, this part is re-did and updated all into Table~\ref{tab:root-causes}}issues in which essential steps or states in the login flow are missing.
For example, issue \#4462 from NextCloud~\cite{nextcloud-android}, which supports user login with multiple accounts, demonstrated that users could not add a new account without replacing the previous one. This occurred because app never updated the boolean variable  \texttt{isFirstRun}, a flag to identify whether it's the first time to log into the app.
As a result, the app mistakenly believed that it was always the first run, and consequently, the new account information will overwrite the previous logged-in information.

\paragraph{Incorrect Login Flow}
This category involved 48 instances in which an app transits to a state or branch that does not align with expected behavior, leading to an incorrect login flow.
Incorrect app transition implementations often cause these issues.
A relevant example is issue \#219 from Home Assistant~\cite{home-assistant-android}: users encounter a white screen after logging in. This problem stems from the application's handling of \texttt{PREF\_REMOTE\_URL}, which stores the redirecting URL after use log in. This issue occurs when a null value is passed to \texttt{PREF\_REMOTE\_URL} (e.g. when no URL was input from a user). Instead of using the existing default URL, the null value overrides the default one and leads to a white screen, representing a wrong consequence of the state transition.

\subsubsection{API Misuse}
The misuse of APIs, particularly those related to authentication, such as Google or Facebook login APIs, emerged as a significant root cause. This category includes 63 identified issues where developers often use login-related APIs or their return values incorrectly.
For instance, issue \#8525 in NextCloud showed the app crashing when attempting to login when trying to access a null token.
The issue was induced by using a wrong API (\texttt{peekAuthToken}) from Android \texttt{AccountManager}.
The developers overlooked that \texttt{peekAuthToken} will immediately return null if no token is cached for the app.
To fix the issue, \texttt{blockingGetAuthToken} is used instead to ensure that the application does not proceed without a valid token.

\subsubsection{Null Object Check Omission}
The omission of null object checks accounts for 39 issues where developers failed to verify whether an object was null before accessing it. For example, in NextCloud~\cite{nextcloud-android}, issue \#9971 reported a crash when sharing content to NextCloud without a prior login. Previously, the app directly called \texttt{getUser()}, but this method did not include a null check. The crash was resolved by updating the code first to check \texttt{getUser().isPresent()} before proceeding, ensuring that user information is available.

\subsubsection{Improper Encoding Algorithm}
Incorrect application of encoding algorithms, such as URL or Base64 encoding, induced 23 issues in our dataset. Problems often arose when an app did not adequately encode special characters in user credentials, leading to authentication errors. For example, in ownCloud~\cite{owncloud}, issue~\#2451 involved a login failure when the password contained the special character ``§". Initially, the code did not specify the character encoding when creating basic credentials, using \texttt{Credentials.basic(mUsername, mPassword)}.
As a result, this character cannot be encoded.
The issue was fixed by modifying the code to explicitly use UTF-8 encoding, changing to \texttt{Credentials.basic(mUsername, mPassword, Util.UTF\_8)}, thereby ensuring that all characters can be correctly handled in the credentials.

\subsubsection{Wrong Environment Dependency}
Finally, 22 issues were attributed to incorrect environment dependencies, where login functionalities failed due to being executed in unsuitable or misconfigured environments, such as specific Android versions. For example, in Thunderbird issue \#2146~\cite{thunderbird-android}, users could not log in and encountered an \texttt{SSLHandshakeException} after updating to Android 7.0. The developer explained that this was expected, as SSL cryptography in Android 7.0 is ``broken'' due to the absence of certain widely used elliptic curves. They advised users to upgrade to Android 7.1.1 or higher to resolve the issue.


\begin{tcolorbox}[left=3pt,right=3pt,top=1pt,bottom=1pt]
    \textbf{Answer to RQ1:} 
\textit{We identified five categories of common root causes for login issues in Android apps. The most prevalent root cause is Login Flow Errors, where the login flows in the apps are broken due to missing steps, incorrect state transitions, improper error handling, or interference with other Android-native state machines such as the Activity lifecycles.
}
\end{tcolorbox}
\subsection{RQ2: Symptoms}
Understanding the symptoms of login issues can help developers design effective test oracles.
In investigating login issues in Android apps, we carefully examined the descriptions of issues, developer responses, and the differences between the buggy and fixed code snippets. We identified seven major categories of symptoms manifested by login issues. The results are shown in Table~\ref{tab:symptoms}.

\begin{table}
  \centering
  \caption{Summary of Login Issue Symptoms}
  \label{tab:symptoms}
  \begin{tabular}{@{}c l r@{}} 
    \toprule
    \textbf{No.} & \textbf{Symptoms} & \textbf{\#Issues} \\ 
    \midrule
    1 & Crash & 85 \\
    2 & Login Delays or Timeout & 57 \\
    3 & Incorrect Navigation Flow & 51 \\
    4 & Account Management Function Break & 48 \\
    5 & Credential Rejection & 41\\
    6 & User Interface Error & 40\\
    7 & Inaccurate Error Message Displayed & 39 \\
    \bottomrule
  \end{tabular}
\end{table}

\subsubsection{Crash}
85 of our studied issues are crashes. They frequently occur due to unhandled exceptions within the login process, significantly disrupting user experience and undermining the application's reliability. For example, issue \#109 of Home Assistant~\cite{home-assistant-android} identified that the app crashed when users tried to paste their username or password into the login screen. This issue was caused by conflicts with the Localise SDK, which interfered with clipboard functionality, leading to unhandled exceptions.

\subsubsection{Login Delay or Timeout}
We documented 57 instances where users experienced significant delays or timeouts during the login process. These issues predominantly involve excessive waiting time or complete failures to log in, which can significantly hinder user experience and service access. For instance, issue \#10043 from WooCommerce~\cite{woocommerce-android} mentions a scenario where users attempting to login Jetpack~\cite{Jetpack} using usernames and passwords encounter a premature dismissal of the WebView before they can approve the connection. This abrupt interruption prevents successful authentication, leading to timeouts and disrupting the overall login process.

\subsubsection{Incorrect Navigation Flow}\label{IncorrectNavigationFlow}
51 issues induced incorrect navigation flows, where users are not directed to the correct screen after successfully logging in. For example, The \textit{Open Event Attendee Android} application~\cite{open-event-attendee-android}, developed by FOSSASIA~\cite{fossasia}, is designed to facilitate event participation by allowing users to view event details, book tickets, and more, directly from their mobile devices. Issue \#1762 occurs when users cannot resume their previous activities, such as viewing tickets, after successfully logging in. This problem arises because the application fails to keep track of the user's progress. To remedy this, developers need to ensure that the app retains the user's last activity state through the login process, allowing them to return to their actions seamlessly after authentication.

\subsubsection{Account Management Function Break}
We identified 48 issues as account management function breaks. These issues are failures in account management functionalities such as adding or managing multiple accounts. An example is discussed in Section~\ref{Missingloginflow}, where a user is unable to add a new account.

\subsubsection{Credential Rejection}\label{CredentialRecognitionErrors}
41 issues were reported as credential rejections where valid login credentials were entered but mistakenly denied by the apps.
For example, issue~\#10170 from NextCloud~\cite{nextcloud-android} demonstrates that the login process fails when users try to login with the credentials containing whitespaces.
This was induced by the wrong encoding method of the app. 

\subsubsection{User Interface Error}
We identified 40 issues related to User Interface (UI) errors. UI errors include interface freezes that prevent user interaction and malfunctioning UI elements, collectively blocking the login process. For example, issue \#1920 of \textit{Open Event Attendee Android}~\cite{open-event-attendee-android} exhibits UI freezing, where the app becomes non-responsive to user input following Google login initiation. Similarly, issue \#1104 of PocketHub~\cite{PocketHub} demonstrates UI element malfunction through an inactive authorization control, where the authorization button remains disabled during credential submission.

\subsubsection{Inaccurate Error Message Displayed}\label{InaccurateErrorMessageDisplayed}
39 issues were identified where users receive inaccurate or misleading error messages. These incorrect messages often do not reflect the underlying problem accurately, inducing user confusion and frustration. For example, in LibreTube~\cite{LibreTube}, an open-source video streaming app, issue \#3104 documented three scenarios displaying incorrect error messages when handling authentication errors.
Scenario (A) involved a user incorrectly clicking the login button without an account, leading to an HTTP 401 displayed. Instead, a user-friendly warning should have been displayed to explain the problem. In scenario (B), errors occur when users attempt to re-access the already deleted accounts. However, the error message indicated that this was a network error. Scenario (C) described an HTTP 401 displayed when a user attempted to re-register with the same username and password.
The core issue of all these three scenarios was the absence of appropriate error messages corresponding to underlying problems.

\begin{figure}[ht!]
  \centering
\includegraphics[width=0.95\linewidth]{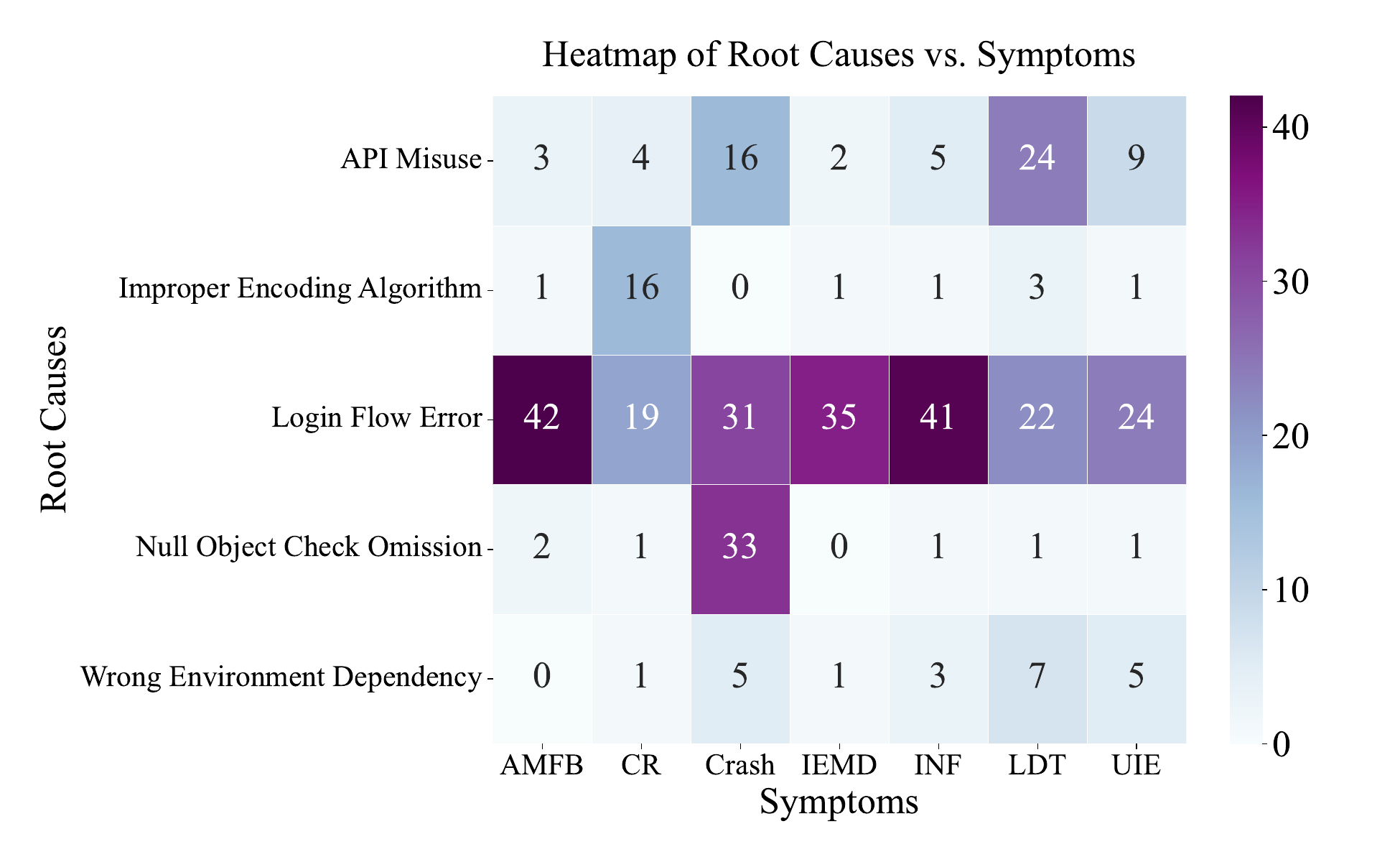}
  \caption{Heatmap between Root Causes and Symptoms.  Abbreviations: AMFB (Account Management Function Break), Crash (Crash), IEMD (Inaccurate Error Message Displayed), INF (Incorrect Navigation Flow), UIE(User Interface Error), CR(Credential Rejection), LDT(Login Delays or Timeout).}
  \label{fig:btRCvsSymp}
\end{figure}

To summarize, despite the fact that crashes are the most prevalent symptoms, they only account for 23.54\% of the studied issues. In other words, around 80\% of the login issues cannot be detected by simply using crashes as the test oracle.
This indicates the need for effective oracles specific to the login issues.
Our symptom categories can serve as the guideline for designing such oracles.
Figure~\ref{fig:btRCvsSymp} depicts a heatmap between root causes and symptoms analysis.
Notably, in every category of symptoms, login flow errors are always the leading cause, demonstrating their significance in ensuring the reliability of the login processes in Android apps. 

\begin{tcolorbox}[left=3pt,right=3pt,top=1pt,bottom=1pt]
    \textbf{Answer to RQ2:} 
\textit{
Our analysis results indicate that login issues can exhibit a variety of symptoms.
Around 80\% of the login issues do not exhibit crashes when triggered. This highlights the importance of designing effective test oracles to expose login issues. Our categories can serve as the guidelines for designing such test oracles.
}
\end{tcolorbox}
\subsection{RQ3:Trigger Conditions of Login Issues}

Understanding the trigger conditions of login issues is crucial for designing effective test cases to expose login issues. In RQ3, we aim to identify the trigger conditions of login issues by analyzing the bug reports, code revisions, and discussions between users and developers within our dataset.

Unlike root causes and symptoms,a single issue can require multiple conditions to trigger.
As a result, the categories of trigger conditions for an issue are not mutual exclusive.
This approach resulted in nine common trigger conditions, with a summary shown in Table~\ref{tab:trigger-conditions}.

\begin{table}
  \centering
  \caption{Summary of Trigger Conditions}
  \label{tab:trigger-conditions}
  \begin{tabular}{@{}c l r@{}} 
    \toprule
    \textbf{No.} & \textbf{Trigger Conditions} & \textbf{\#Issues} \\ 
    \midrule
    1 & Specific Login Approach & 122 \\
    2 & Account Management Operation & 93 \\
    3 & Specific User Input & 86 \\
    4 & No Specific Trigger Conditions & 38 \\
    5 & Communication Failure with the Cloud & 37 \\
    6 & Specific Device & 30 \\
    7 & Setting Change & 20 \\
    8 & Orientation Change & 18 \\
    9 & User Re-Login & 15 \\
    
    \bottomrule
  \end{tabular}
\end{table}

\subsubsection{Specific Login Approach}
The most frequent trigger condition was associated with specific login approaches, accounting for 122 issues.
Issues in this category can only be triggered under specific login approaches (e.g., using specific third-party login services or using MFA).
For example, issue \#4393 in Tusky~\cite{tusky} described how selecting the Akkoma account type for login led to an app crash. This example highlights the necessity of comprehensive test coverage for each supported login approach to prevent functional disruptions across different authentication scenarios.

\subsubsection{Account Management Operation}
Account management operations triggered 93 issues, where account management actions such as account creation, deletion, and account switch triggered problems. For example, issue \#6481 from NextCloud~\cite{nextcloud-android} is triggered when users try to add a new account to the app. This suggests the importance of covering the different account management operations in testing login issues.

\subsubsection{Specific User Input}
86 issues required specific inputs to trigger. These issues often arose from scenarios where special characters or unexpected data types in login forms. The issue we discussed in Section~\ref{CredentialRecognitionErrors} is an example. The developers may not anticipate all the variants of the user input, and such issues are triggered. Consequently, test cases including special  characters or data types are necessary for login issue testing.

\subsubsection{No Specific Trigger Conditions}
Finally, there are 38 issues that do not require any specific trigger conditions.
These issues can be triggered by any user login attempts.

\subsubsection{Communication Failure with the Cloud}
Communication failures with cloud services triggered 37 issues.
These issues often occur during synchronization processes or when fetching authentication tokens. This trigger condition indicates the dependency on external cloud infrastructures and the need for effective handling of network-related uncertainties.

\subsubsection{Specific Device}
Device-specific conditions accounted for 30 issues involving various hardware configurations or operating system versions. These issues are essentially compatibility issues occurring in the login process~\cite{chen2024demystifying,7582761}. This highlights the challenges in developing robust applications across diverse hardware and software ecosystems. To avoid issues triggered by device-specific conditions, developers must execute all test cases on a range of devices with different operating system versions before releasing updates.

\subsubsection{Setting Changes}
20 issues were triggered by changes in the app settings, including privacy setting updates or security configurations that inadvertently impacted the login functionality.
This calls for testing the login functionalities under different settings.

\subsubsection{Orientation Change}
18 issues required device orientation changes to trigger, illustrating problems that occur when apps fail to handle changes in device orientation (i.e., changes in Activity lifecycles) during login processes. This condition highlights the need for responsive design strategies that ensure seamless user experiences regardless of device orientation.
It also calls for the developers to properly handle the interference between the login flow and the Activity lifecycles.

\subsubsection{User Re-Login}
15 issues were triggered only when users make a second attempt to log into the app.
This includes situations where users aim to change the account or add a new account.
This trigger condition points to the necessity of designing test cases to re-login in a short period.


\begin{figure}[ht!]
  \centering
  \includegraphics[width=0.9\linewidth]{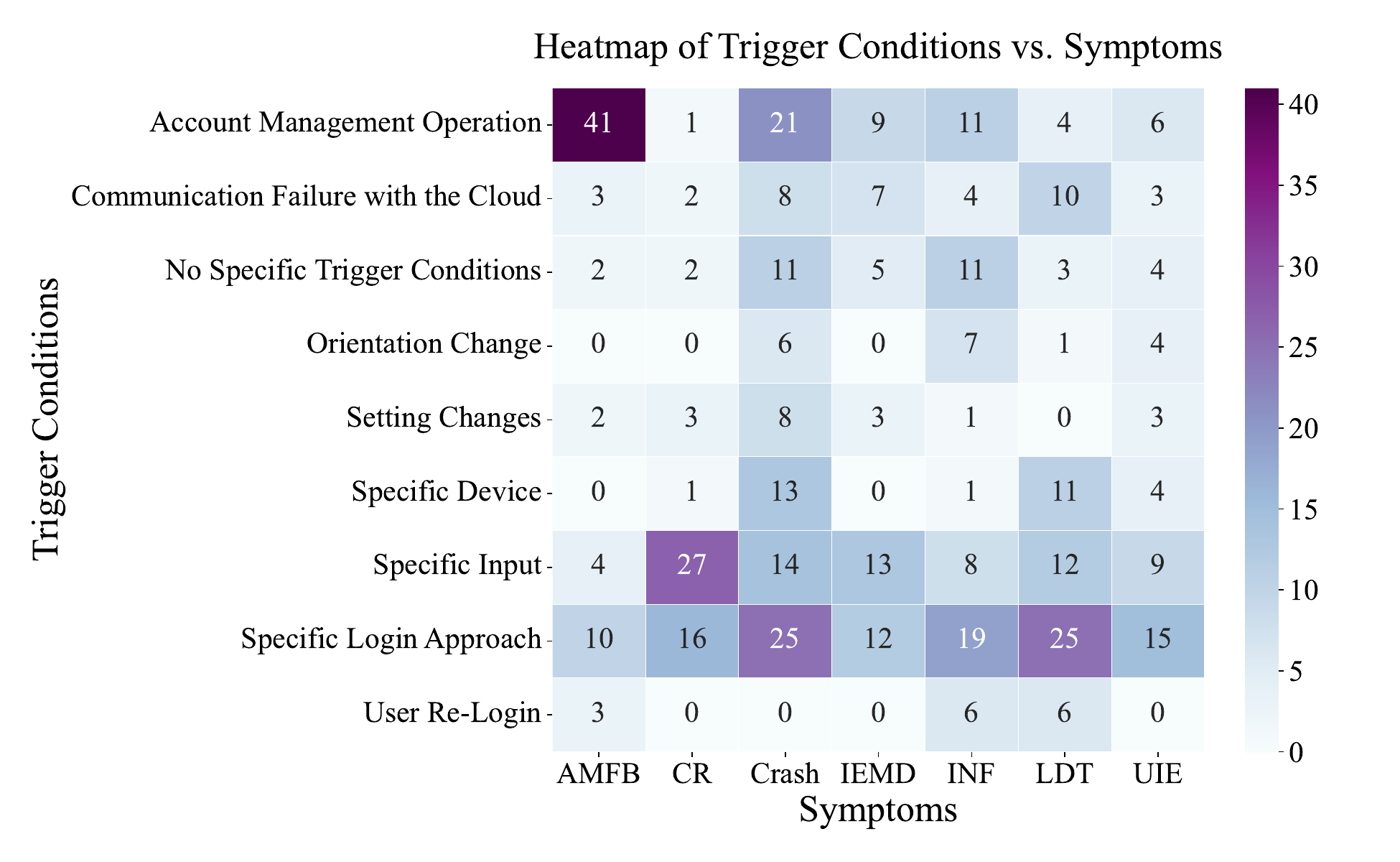}
  \caption{Heatmap between Trigger Conditions and Symptoms. Abbreviations: AMFB (Account Management Function Break),Crash (Crash), IEMD (Inaccurate Error Message Displayed), INF (Incorrect Navigation Flow), UIE(User Interface Error), CR(Credential Rejection), LDT(Login Delays or Timeout).}
  \label{fig:btTCvsSymp}
\end{figure}


Our results show that the majority (89.7\%) of the login issues require complicated conditions to trigger.
This highlights the complexity in testing login issues.
Furthermore, the heatmap presented in Figure~\ref{fig:btTCvsSymp} correlates trigger conditions with symptoms, guiding developers in designing effective test cases. For instance, the heatmap suggests that login delays or timeouts can be tested through a combination of four main conditions: specific login approaches, specific input, specific device and communicating with could failures. Developers are advised to create test cases that incorporate special character inputs across all designated login approaches and execute these under diverse network scenarios to simulate and address potential failures.
\begin{tcolorbox}[left=3pt,right=3pt,top=1pt,bottom=1pt]
    \textbf{Answer to RQ3:} \textit{The majority (89.7\%) of the studied login issues require complicated conditions to trigger, highlighting the difficulties in triggering and testing login issues in Android apps. Our analysis identified nine common trigger conditions. The predominant conditions include specific login approaches, account management operations and specific inputs. Our summarized trigger conditions can guide developers to better test their app login processes.
}
\end{tcolorbox}

\subsection{Implications}

The findings from our study emphasize the importance of login flow errors in the context of login issues within Android apps. Login flow errors are the primary root causes for our studied issues, accounting for 59.28\%. 
In this section, we further investigate the impact of login flow errors by correlating them with the issue symptoms and trigger conditions.
\subsubsection{Root Causes and Symptoms}
As shown in Figure~\ref{fig:btRCvsSymp}, unlike other root causes that typically correspond to specific symptoms (e.g., null object check omission always induce crashes), login flow errors can trigger every type of symptom identified in our study. This broad impact suggests that issues in login flows can manifest in various detrimental ways.
Here we use examples to illustrate how login flow errors induced two types of symptoms as examples:

\blacknumber{1}\textbf{Incorrect Navigation Flow:} Login flow errors induced 41 incidents of incorrect navigation flow.  The issue mentioned in Section~\ref{IncorrectNavigationFlow} reports that users are unable to resume their previous activities after logging in.  The root cause of this issue is \textit{Interference with Other State Machines}. The app did not use \texttt{savedInstanceState} to save the activity before starting to login, resulting in information loss. To fix this issue, the developers implemented a solution where the user's current activity state is saved before authentication. Upon successful login, the application restores the previously saved state, allowing users to seamlessly continue from where they left off.

\blacknumber{2}\textbf{Inaccurate Error Messages:} 35 inaccurate error message issues were induced by login flow errors.
In Section~\ref{InaccurateErrorMessageDisplayed}, we discussed an issue in LibreTube where error messages do not match the underlying root causes.
The root cause of this issue is \textit{Improper Error State Handling}. 
The login flow transits to incorrect error states and the app did not give precise error messages in the the error-handling logic.
  
\subsubsection{Root Causes and Trigger Conditions}
As shown in Figure~\ref{fig:btTCvsSymp}, login flow errors can be triggered by all kinds of conditions.
In addition, they often require a confluence of multiple trigger conditions to manifest, suggesting their complex nature in issue triggering. Here are two examples where multiple conditions are required simultaneously  to trigger login flow issues:

\blacknumber{1}Thunderbird~\cite{thunderbird-android} is an open-source email client developed by Mozilla that supports multiple email accounts, management, and protocols, making it popular for personal and professional communication.
In email systems, the port number is critical as it dictates the communication endpoint for a server. Different email services use specific ports to handle various email protocols.
For instance, IMAP typically uses port 143 or the encrypted 993.
The root cause of issue \#5088 
in Thunderbird~\cite{ThunderbirdIssue} is identified as \textit{Missing Login Flow}. The developers missed the process in the login flow to update the port number according to the protocol set by the new account when users switch between email accounts. 
This problem can be triggered when three conditions are satisfied: 1) the user adds a new account (\textit{Account Management Operation}), 2) initially sets up the account using the IMAP protocol (\textit{Specific Login Approach}), and 3) inputs a specific server address (\textit{Specific input}). These steps inadvertently lead to a situation where the port number from the previous account setup persists, causing connection issues or incorrect email data retrieval. 

\blacknumber{2}AntennaPod~\cite{AntennaPod} is an open-source podcast manager that allows users to sync their podcast data across devices using various login and synchronization methods, including NextCloud~\cite{nextcloud-android}. 
Issue \#5841 has been documented within AntennaPod's integration with NextCloud~\cite{AntennaPodissue}. 
This integration requires users to provide a server URL for syncing data.
In this issue, despite users entering the correct server URL, they encounter persistent login failures. The root cause is identified as \textit{Missing Login Flow} because the developers did not design the login flow for the input URLs with different certificates.  
This problem is triggered when three conditions are satisfied: 1)the device must have \texttt{Gpoddersync} installed (\textit{Specific Device}), 2) NextCloud selected as the sync and login approach (\textit{Specific Login Approach}), and 3) the entered URL must support HTTPS certificates (\textit{Specific Input}). The necessity for an HTTPS-supportive URL suggests that security protocols, particularly those related to SSL/TLS certificate verification, may be obstructing the login process. This issue can lead to unsuccessful authentication and synchronization, severely affecting the user experience and functionality of the app.

In conclusion, our study results reveal the pivotal role of login flow errors in compromising the functionality and security during the login process of Android applications. 
\subsubsection{Findings}
To summarize, the login flow errors can cause a wide range of symptoms, from login delays or timeouts to system crashes, and require the intersection of multiple conditions to trigger.
This indicates that further research should focus on login flow errors when testing for login issues in Android apps. First, to tackle these issues, it is essential to understand the states and transitions involved in the login flow and develop models that accurately represent these dynamics.
Our proposed state machine (Figure~\ref{fig:statemachine}) is a first attempt to achieve this goal.
Second, our results assist developers in identifying the causes of problems and designing targeted test cases. For instance, if users report an Incorrect Navigation Flow issue, developers should focus on login flow errors and verify the logical sequence of state transitions.
Additionally, developers can design tests that include special character inputs for all supported login methods and incorporate device orientation changes to assess the Incorrect Navigation Flow.
Third, our taxonomy of issue symptoms can help define effective test oracles. For instance, we identified 39 issues of “Inaccurate Error Message Displayed”. This indicates that developers can design test oracles to verify whether error messages accurately correspond to actual failures like invalid credentials or expired sessions.
Such test cases not only confirms issue fidelity for users but also ensures that the application's logic proactively manages these errors, thus improving both the reliability and user experience of the apps' login process. 

\section{Related Works}
\subsection{Empirical Study On Other Issues in Android}
Several studies have explored the challenges and bugs in Android apps, focusing on issues related to user accessibility~\cite{9525343}, GUI~\cite{7965405}, and configuration~\cite{jha2019empirical}. Additionally, numerous investigations into Android testing have been conducted. Fabiano et al.~\cite{pecorelli2022software}, Lin et al.~\cite{10.1145/3324884.3416623}, and Xiong et al.~\cite{10.1145/3597926.3598138} provide empirical studies on functional bugs in Android apps, though their focus remains predominantly on common bugs rather than specific login-related issues which critically impact user login. Ardito et al.~\cite{9217524} present an automated test selection framework for Android apps based on APK and activity classification, while Yan et al.~\cite{10.1145/3377811.3380347} propose a multiple-entry testing framework by constructing activity-launching contexts. Furthermore, Bose et al.~\cite{10172528} propose a systematic callback exploration framework for Android testing. Despite the advancements in testing techniques, these tools often do not fully address the complex features of login processes, such as authentication and session management, which are critical areas that our study specifically targets. 

These studies, while comprehensive in their scope, often overlook the nuanced complexities of login mechanisms and their repercussions on security and usability. Our research aims to fill this gap by focusing specifically on the systematic analysis of login issues, offering new insights into their causes, symptoms, and triggers that can inform more effective testing and development practices for login processes.


\subsection{Studies On Android Authentication Problems}
Several studies have also focused on the authentication processes in Android apps, yet none have conducted a comprehensive and systematic investigation of the entire login process.
Wan et al.~\cite{8919059} attempt to test user interfaces (UIs) within mobile applications, including those for login screens. However, their tools are unable to adequately handle UIs that require user input, such as entering usernames and passwords, posing significant challenges in fully automating the testing process for such critical components. Song et al.~\cite{9401988} developed VPDroid, a platform designed to test the automatic login features of Android apps. While their approach effectively evaluates the security of apps that rely on device consistency checks for auto-login functionalities, it does not extend to other login scenarios such as Google login or server address authentication, which are covered in our study. Several studies focus on specific login approaches, Ma et al.~\cite{10.1145/3359789.3359828} conducted an empirical study on one-time password authentication in Android.  Jannett et al.~\cite{10628996} developed a tool called SoK focused on Single Sign-on in Android. Shi et al.~\cite{10.1145/3321705.3329801} propose MoSSOT, a black box tester for OAuth. In addition, Tamjid et al.~\cite{8952200} conducted an empirical study on the usage of OAuth APIs and their implications for mobile security, leading to the development of OAUTHLINT, a tool that employs query-driven static analysis to identify vulnerabilities in OAuth implementations on the Google Play marketplace. While these studies offer significant insights into various aspects of login and authentication mechanisms within Android apps, they typically focus on specific components or security aspects. This narrow focus often bypasses the comprehensive challenges presented by the entire login process, particularly in how different authentication methods integrate and operate under diverse conditions. Our research seeks to bridge this gap by conducting a systematic and holistic analysis of the login process across authentication scenarios, not just limited to OAuth or single sign-on systems. By doing so, we aim to provide a more complete understanding of the functionality issues that can arise during user login in Android apps.  
\section{Threats to Validity}

\textbf{Data Source and Diversity}: While we have exclusively sourced our repositories from GitHub and all are open-sourced, this may limit the generalizability of our findings. However, our dataset includes a diverse array of app categories, enhancing its representativeness. For example, we have categorized apps into productivity tools like \textit{Orgzly}~\cite{orgzly} and \textit{nextcloud}~\cite{nextcloud-android}, social media apps such as \textit{Duckduckgo}~\cite{duckduckgo} and \textit{Facebook Android SDK}~\cite{facebook-android-sdk}, multimedia applications like \textit{Pocket Casts}~\cite{pocket-casts-android} and \textit{AntennaPod}~\cite{AntennaPod}, and security-focused apps including \textit{Password Store}~\cite{Password-Store} and \textit{Aegis}~\cite{Aegis}. According to Table~\ref{fig:data selection}, the selected repositories are highly popular, evidenced by high numbers of GitHub stars and extensive download numbers on Google Play. These repositories are also well-maintained, further attesting to the reliability of our data.

\textbf{Classification Subjectivity}: The categorization of apps was performed manually, which could introduce bias. To mitigate this, we employed an open coding approach, with discrepancies resolved through discussions among the research team. The inter-rater reliability, measured by Cohen’s kappa, was 0.81, indicating substantial agreement and reliability in our classification method.

\textbf{Platform Specificity}: Our study focuses solely on Android apps. While this provides in-depth insights into state management issues within this platform, the findings may also apply to other platforms due to the generic nature of login functionalities. This suggests that the problems identified, while analyzed in the context of Android, could potentially manifest in similar ways on other operating systems, thus providing a broader relevance to our conclusions.


\section{Conclusion and Future Works}
In this paper, we conduct an empirical study of 361 login issues from 44 widely-used open-source Android apps. Our research discloses the common root causes and symptoms of Android login issues and explores the various factors that trigger these problems. The insights gained from our study are crucial for guiding future research into testing and diagnosing Android login issues. Moving forward, we aim to expand our dataset for a larger-scale empirical analysis. Building on our current findings, we also plan to develop a dynamic analysis tool that models the login processes using state-machines. In addition, we plan to generate both test oracles and test cases, aligning with the patterns identified in our study.
\section{acknowledgment}
We thank the anonymous reviewers for their insightful comments on the paper.
This work was supported by Natural Sciences and Engineering Research Council of
Canada Discovery Grant (Grant No. RGPIN-2022-03744 and Grant No.
DGECR-2022-00378), FRQNT/NSERC NOVA program (Grant No. 2024-NOVA-346499) and National Natural Science Foundation of China (Grant No. 62372219).

\balance
\bibliography{references}

\end{document}